\documentclass[%
reprint,
amsmath,amssymb,
aps,
prl
]{revtex4-2}

\usepackage{graphicx}% Include figure files
\usepackage{dcolumn}% Align table columns on decimal point
\usepackage{bm}% bold math
\usepackage{hyperref}% add hypertext capabilities
\usepackage{color}

\begin{document}

\preprint{APS/123-QED}

\title{Stabilization of three-body resonances to bound states in a continuum}

\author{Lucas Happ}
\email{lucas.happ@riken.jp}
\affiliation{Few-body Systems in Physics Laboratory, RIKEN Nishina Center, Wak\={o}, Saitama 351-0198, Japan}%Lines break automatically or can be forced with \\
\author{Pascal Naidon}%
\affiliation{Few-body Systems in Physics Laboratory, RIKEN Nishina Center, Wak\={o}, Saitama 351-0198, Japan}

\date{\today}% It is always \today, today,

\begin{abstract}
Three-body resonances are ubiquitous in quantum few-body physics and are characterized by a finite lifetime before decaying into continuum states of their composing subsystems. In this work we present a theoretical study on the possibility to stabilize three-body resonances to so-called bound states in a continuum: resonances with vanishing width that do not decay. Within a two-channel approach we unveil the underlying mechanism and show how the lifetime can be made infinitely long by a continuous tuning of system parameters. The validity of our theory is illustrated in two different examples: a mass-imbalanced system in one dimension and a system of three identical bosons in three dimensions, relevant to Efimov physics. Crucially, for the latter we find that one of the parameters that can be tuned to achieve a three-body bound state in a continuum is an external magnetic field, a common tunable variable in cold-atom experiments. Due to the generality of this stabilization effect, it is expected to be applicable to a wide range of unstable few-body systems, opening new perspectives for fundamental studies as well as technical applications.
\end{abstract}

\maketitle

%\tableofcontents

\section{Introduction}\label{sec:introduction}
Few-body resonances are crucial in many fields of physics, spanning from nuclear physics where the famous Hoyle state~\cite{hoyle1954} is essential for nucleosynthesis, hadron physics where hadronic molecules~\cite{hyodo2013,guo2018} can describe exotic hadrons, ultracold atoms experiments where triatomic Efimov states have been observed~\cite{kraemer2006,knoop2009} to few-body analog systems of excitons in semiconductors~\cite{kazimierczuk2014,belov2024}. Few-body resonances are metastable states that can spontaneously decay into one or several configurations of their subsystems~\cite{kukulin1989,moiseyev1998}. The arguably most remarkable few-body states, Efimov trimers~\cite{efimov1970,efimov1973,naidon2017}, are in fact usually resonances and not true bound states. Indeed, the first observation~\cite{kraemer2006} of Efimov states, performed in a system of ultracold atoms, was based on measurements of the three-body recombination rate, a phenomenon only possible due to the existence of deeply bound two-body states and decay into the associated continua~\cite{esry1999,dincao2004}. In contrast to the majority of works on Efimov trimers, which focus on their resonance position, there are comparatively few studies on their resonance width~\cite{penkov1999,nielsen2002,pricoupenko2010a,happ2024}. However, advancements in spectroscopy~\cite{lompe2010,chuang2025} and interferometric implementations~\cite{yudkin2024} show great prospect for near-future measurements of trimer lifetimes. Besides the Efimov scenario, three-body resonances and their stability are also relevant to collisions between ultracold atoms and molecules, a topic that has recently attracted considerable experimental and theoretical interest~\cite{knoop2011,wang2021a,nichols2022,son2022,bause2023,park2023,cao2024}.

The central property that distinguishes resonances from bound states is their finite width and associated decay into one or possibly many continuum states. However, there can be exceptional scenarios when a resonance defies decay and remains stable, despite being embedded in a continuum. Such states are called bound states in a continuum (BIC)~\cite{hsu2016} and have been proposed and studied over the course of nearly a century, from early theoretical works~\cite{vonNeumann1929,stillinger1975,friedrich1985}, to experimental realizations, most notably in photonics~\cite{marinica2008,pankin2020}, with more recent theoretical proposals in cold-atom systems~\cite{deb2014,chilcott2024}. Although no such BIC has been demonstrated for few-body resonances yet, Ref.~\cite{happ2024} presented a strong increase of three-body resonance lifetimes for specific mass ratios in a heteronuclear atomic three-body system. However, this study could neither reveal whether the state is actually stable, nor its stabilization mechanism.

In this article we unveil the underlying physical mechanism that can stabilize three-body resonances to BICs by employing a two-channel Feshbach-like model~\cite{feshbach1958,fano1961}. We find that the effect can be traced back to a transition element which can be tuned to become zero. Critically, our two-channel model reveals that stabilization can be achieved for a single resonance in a single continuum. No interference of several resonances is required. Hence, we can classify them as so-called \textit{single-resonance parametric BICs}~\cite{hsu2016}. The key tuning parameter turns out to be the relative momentum between outgoing subsystems which can indeed be tuned via the mass ratio, as well as other parameters. We demonstrate three-body BICs in two example systems: (i) a mass-imbalanced system in one dimension, and (ii) three identical bosons in three dimensions, relevant to the Efimov scenario. Crucially, for the latter we find that stabilization of a three-body resonance to a BIC can occur via variation of an external magnetic field, a parameter commonly tuned in cold-atom experiments~\cite{inouye1998a,chin2010}.

Stabilizing Efimov states has been a long-standing challenge \cite{laird2018} for which our demonstration of three-body BICs provides a promising solution. It opens a path towards realization of quantum systems with strong three-body interactions~\cite{bulgac2002}, trimer condensates~\cite{braaten2003,musolino2022}, Efimov liquid~\cite{piatecki2014}, and allows studying effective atom-dimer potential surfaces and three-body interactions in much greater detail~\cite{yang2022a,endo2025}. If experimentally tunable parameters can also be found in other three-body systems, e.g. in nuclear or hadron physics, they could serve as a tool to extend their lifetime, opening new opportunities to study their properties in more detail. Beyond the fundamental interest of studying trimer BICs, the region around a point of stability offers the ability to manipulate a trimer's lifetime over several orders of magnitude, which can be used for engineering purposes. For instance, long-lived trimers are relevant to the realization of quantum simulators~\cite{rapp2007,bloch2012}.

\section{Results}
\subsection{Two-channel description}\label{sec:generaltheory}
To unveil the mechanism responsible for the formation of BICs in quantum few-body systems, we introduce a two-channel description
\begin{equation} \label{eq:2chnl}
    H = \begin{pmatrix} H_{\mathrm{o},\mathrm{o}} & H_{\mathrm{o},\mathrm{c}} \\ H_{\mathrm{c},\mathrm{o}} & H_{\mathrm{c},\mathrm{c}} \end{pmatrix}
\end{equation}
of the total few-body Hamiltonian, where the subscripts $\mathrm{o}$ and $\mathrm{c}$ denote the open and closed channels, respectively. The bare states in each channel are defined from the uncoupled equations
\begin{align}\label{eq:bareequations}
(H_{\mathrm{o},\mathrm{o}}-E_{\mathrm{o},m})|\Phi_{\mathrm{o},m}\rangle =&~ 0 \nonumber \\
(H_{\mathrm{c},\mathrm{c}}-E_{\mathrm{c},n})|\Phi_{\mathrm{c},n}\rangle =&~ 0
\end{align}
whose Hamiltonians can be expressed as
\begin{equation}
    H_{\mathcal{C},\mathcal{C}} = T + U_\mathcal{C},\qquad \mathcal{C} \in\{\mathrm{o},\mathrm{c}\}
\end{equation}
with the kinetic energy operator $T$, and $U_\mathcal{C}$ denotes the interaction potential in each channel. This effective potential can be obtained e.g. within a Born-Oppenheimer (BO)~\cite{born1927} or hyperspherical~\cite{lin1995,nielsen2001a} approach.

After including the off-diagonal coupling terms $H_{\mathrm{o},\mathrm{c}}$, the previously bare bound states in the closed channel ($E_{\mathrm{c},n} > U_\mathrm{o}(r\to\infty)$) become resonances with finite widths $\Gamma_{n}$ since they can decay into the continuum of the open channel ($U_\mathrm{o}(r\to\infty) < U_\mathrm{c}(r\to\infty)$). For an isolated resonance the width can be calculated via~\cite{kukulin1989,naidon2025}
\begin{equation}\label{eq:gammaisolatedresonance}
\Gamma_{n} = 2\pi \left|\langle\Phi_{\mathrm{o},k}|w_n\rangle\right|^2, \qquad |w_n\rangle = H_{\mathrm{o},\mathrm{c}} |\Phi_{\mathrm{c},n}\rangle.
\end{equation}
where $\Phi_{\mathrm{o},k}$ is the scattering state in the open channel whose scattering energy $E_{\mathrm{o},k}$ matches the energy $E_{\mathrm{c},n}$ of the bound state $\Phi_{\mathrm{c},n}$. We highlight here that this formula for the width involves a single scattering state.

In this article we focus on the qualitative nature of the stabilization effect. Hence, for illustrative purposes we employ the WKB approximation~\cite{wentzel1926,kramers1926,brillouin1926mecanique,l.d.landau1958}
\begin{equation}
    \Phi_{\mathrm{o},k}(r) \propto \frac{1}{\sqrt{k(r)}}\cos\left(\alpha + \int_{}^r \mathrm{d} r' k(r') \right)
\end{equation}
which, while not being required, properly captures the essential features of the continuum state. The WKB phase $\alpha$ ensures the appropriate boundary condition, and the WKB wave number reads
\begin{equation}
    k(r) = \sqrt{\frac{2\mu}{\hbar^2} \Big[E-U_o(r)\Big]}.
\end{equation}
The reduced mass $\mu$ depends on how the channels are defined and on the particular scaling in use~\footnote{In a Born-Oppenheimer description $\mu$ would indicate the heavy-particles reduced mass, whereas in a hyperspherical approach often the $N$-body reduced mass $\mu_N = (\mu_{12} \mu_{12,3}\ldots)^{1/(N-1)}$ is employed, or even $\mu=1$ when the mass is used to rescale the coordinates.}. The overlap integral in Eq.~\eqref{eq:gammaisolatedresonance} can then be expressed as
\begin{align}
    \langle\Phi_{\mathrm{o},k}|w_n\rangle \propto \int \mathrm{d} r\, \frac{\cos\left(\alpha + \int_{}^r \mathrm{d} r' k(r') \right)}{\sqrt{k(r)}} w_n(r).
\end{align}

This describes the integral of an oscillating function times $w_n(r)$. The frequency of the oscillations is mainly governed by $k(r)$. The overlap integral can in general become zero due to cancellations of subsequent positive and negative contributions of the integrand. In that case, the width of the resonance is zero, i.e. the state is a bound state in the continuum. It is however not always guaranteed that total cancellation can occur. Indeed, a simple counterexample would be the integral of a pure sine function times a monotonically decaying exponential.

One can thus see that some non-monotonicity in the function $w_n(r)$ is required. In few-body systems this condition is usually fulfilled since e.g. within the BO or hyperspherical approach, the coupling terms $H_{\mathrm{o},\mathrm{c}}$ contain non-adiabatic derivative operators which ensure a non-monotonic behavior of $w_n(r)$, even if the bare bound state $|\Phi_{\mathrm{c},n}\rangle$ is nodeless.

We have identified that the WKB wave number is a crucial ingredient that determines whether the overlap integral vanishes. For three-body resonances this is closely related to the relative atom-molecule momentum
\begin{equation}\label{eq:krel}
    p_{\mathrm{rel}} = \sqrt{2\mu\left[E^{(3)} - E^{(2)}\right]},\ \ E^{(3)} = E^{(2)} + \frac{ p_{\mathrm{rel}}^2}{2\mu}.
\end{equation}
This momentum is affected by the reduced mass $\mu$ and the energy difference $\Delta E = E^{(3)} - E^{(2)}$, between the resonance position and the binding energy of the deep dimer. Whence, variation of this momentum enables one to tune the value of the overlap integral and therefore allows the width $\Gamma$ of the three-body resonance to vanish. We highlight that this effect originates from a single resonance in a single continuum. Stabilization stems from tuning of the system parameters. No interference of two or more resonances or decay pathways is required.

Needless to say, vanishing transition elements are not unprecedented in physics. A standard example are the selection rules of (dipole-) transition elements in atomic hydrogen, or spin- and angular momentum algebra in nuclear systems. Indeed, in these systems many transition elements vanish due to symmetry arguments or because of addition rules of a discrete parameter. In this sense, one can understand our phenomenon as an analog effect, however based on a continuously tunable parameter. Therefore, our theory predicts BICs of the \textit{single-resonance parametric} type~\cite{hsu2016}.

We note that the presented model describes only the internal lifetime. In an experiment with many particles, collisions with other particles or additional decay channels can further contribute. In the following we demonstrate the generality of our theory in two different scenarios.

\subsection{Illustration in 1D}
For the first example, we consider a three-body system consisting of two identical bosons of mass $M = \beta m$  and a third particle of mass $m$, all confined to one dimension. The mass ratio is given by $\beta$. The identical bosons do not interact directly, while the interspecies interaction is modeled by
\begin{equation}
    V_{BX}(r) = -\left|v_0\right|\exp(-r^2/z_0^2)
\end{equation}
where $r$ is the distance between the interacting particles.

\begin{figure}[htbp]
    \centering
    \includegraphics[width=\columnwidth]{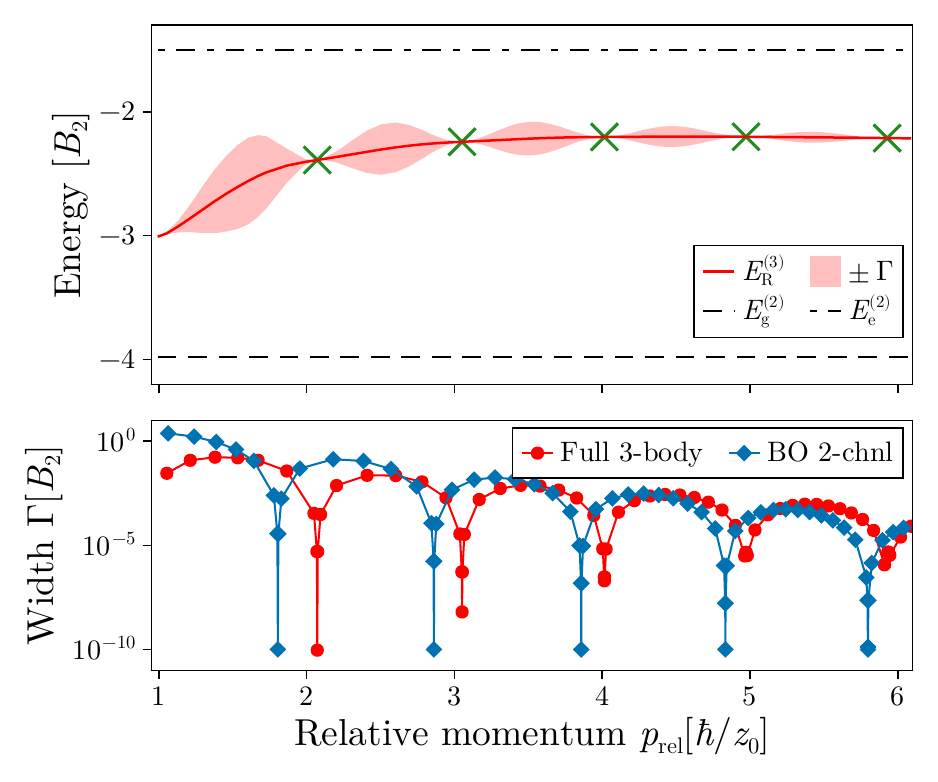}
    \caption{\textbf{Oscillations in the width of three-body resonances in 1D revealing multiple BIC locations.} Position $E_\mathrm{R}^{(3)}$ and width $\Gamma$ of a three-body resonance in a 1D, 2+1 boson system as a function of the relative momentum $p_\mathrm{rel}$, Eq.~\eqref{eq:krel}. All energies are presented in units of the characteristic two-body energy $B_2 = \hbar^2/2\mu_{bx}z_0^2$. Accordingly, $p_{\mathrm{rel}} = \frac{\hbar}{z_0} \, \sqrt{(1+\beta)\big[E^{(3)} - E^{(2)}\big]/B_2}$ is given in units of $\hbar/z_0$, and the momentum is scanned by varying the mass ratio in the range $1/20\leq\beta\leq20$. Upper panel: The three-body resonance (solid red line) is located in the atom-dimer continuum induced by the two-body ground state at energy $E_g^{(2)}$ (dashed black line). The resonance width (shaded area) shows oscillatory behavior with several zero-points (green crosses). For better visibility $\Gamma$ is scaled by the mass ratio $\beta$. Lower panel: resonance width in log-scale  as a function of $p_\mathrm{rel}$, for the full three-body calculation (red line) similar to Ref.~\cite{happ2024} and the BO two-channel model (blue line). Lower bounds of $\Gamma$ are limited by numerical accuracy.}
    \label{fig:1d}
\end{figure}

This three-body problem is then solved in two ways: (i) directly solving the three-body Schr\"odinger equation using the Gaussian expansion method~\cite{hiyama2003,happ2025fewbodytoolkit} with complex scaling~\cite{moiseyev1998}, and (ii) employing the Born-Oppenheimer approximation (see Methods) in which the two deepest potential curves are used to define the open and closed channels of the framework described by Eqs.~\eqref{eq:2chnl}-\eqref{eq:gammaisolatedresonance}.

In Fig.~\ref{fig:1d} we present our results for the three-body resonance position $E_\mathrm{R}^{(3)}$ and its width $\Gamma$ as functions of the relative momentum $p_\mathrm{rel}$. The upper panel (direct three-body calculation) shows that while $E_\mathrm{R}^{(3)}$ varies only slowly, $\Gamma$ exhibits a distinct damped-oscillatory behavior with several zero points. These are the points where the resonance behaves like a stable bound state in the atom–molecule continuum. For this figure, $p_\mathrm{rel}$ is varied via the mass ratio in the range $1/20\leq\beta\leq20$, which is most relevant to ultracold atom experiments. Moreover, all energies are scaled in units of the characteristic two-body energy $B_2 = \hbar^2/2\mu_{bx}z_0^2$, which eliminates the mass-ratio dependence of the two-body binding energies $E_{g,e}^{(2)}$.

The lower panel compares the width $\Gamma$ from the BO two-channel model to the direct three-body calculation. Overall, we find good agreement between the two approaches, giving confidence in our two-channel description. For small $p_\mathrm{rel}$, corresponding to small mass ratios, the discrepancy between the curves is expected from the BO treatment and gradually decreases as $p_\mathrm{rel}$ increases. Although the two-channel description is predominantly qualitative, it nevertheless reproduces the correct number of stabilization points within the studied range, with their locations consistently shifted to smaller $p_\mathrm{rel}$. Since the BO calculation of the width, based on Eq.~\eqref{eq:gammaisolatedresonance}, shows that the overlap integral changes sign due to oscillations in the scattering state, we conclude that the apparent zeros in the exact calculations must also be true BICs, whose nature is thus elucidated as single-resonance parametric BICs.

\subsection{Efimov scenario}

As a second example we consider a different regime where the three-body resonance lies close to the three-body binding threshold. The system consists of three identical bosons of mass $m$ in 3D, making the results relevant to Efimov physics.

\begin{figure}[htbp]
    \centering
    \includegraphics[width=\columnwidth]{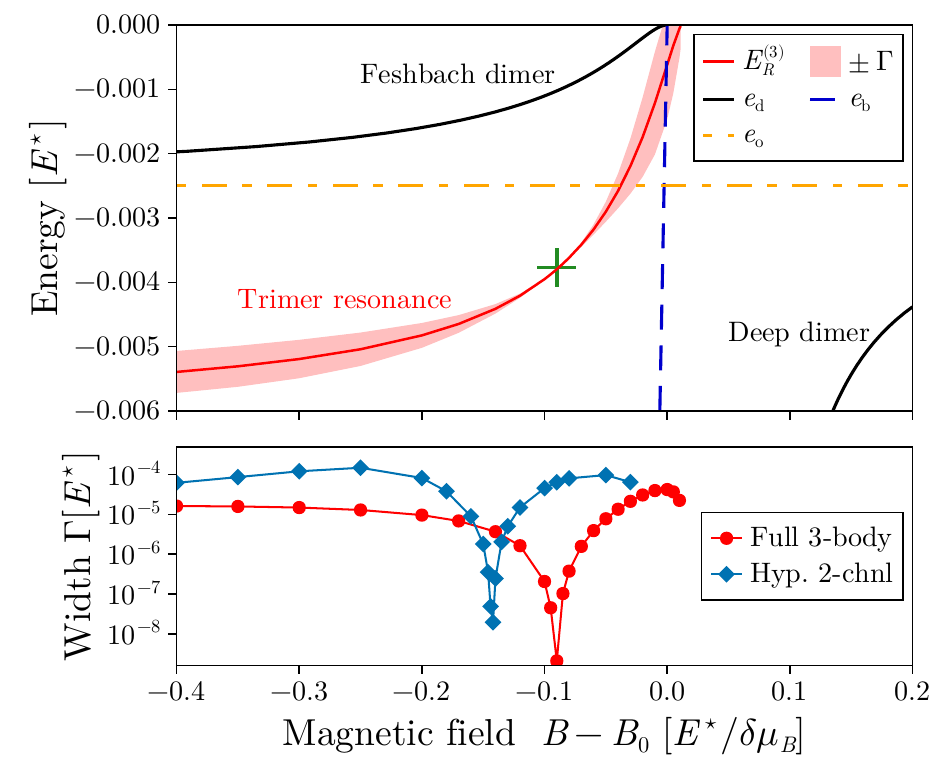}
    \caption{\textbf{Magnetic-field-induced BIC formation in the Efimov scenario.} Position $E_\mathrm{R}^{(3)}$ and width $\Gamma$ of a resonance of three identical bosons in 3D, as a function of the magnetic field $B$. Energies are given in units of $E^\star = \hbar^2/m {R^\star}^2$. Upper panel: The location, where the resonance width vanishes is marked by a green cross. Additionally, three curves characterizing the two-body Feshbach resonance are displayed: the bare bound state energy $e_\mathrm{b}$ (dashed dark blue line), the open-channel bound state energy $e_\mathrm{o}$ (dash-dotted orange line), and the dressed two-body states $e_\mathrm{d}$ (solid black lines). Lower panel: width $\Gamma$ of the three-body resonance in log-scale for the full three-body calculation (red line), and the hyperspherical two-channel model (blue line).}
    \label{fig:3d1}
\end{figure}

Such Efimov trimer resonances are observed in ultracold bosonic atom experiments close to a magnetic Feshbach resonance~\cite{tiesinga1993,inouye1998a,chin2010}, i.e. when the intensity $B$ of an externally applied magnetic field approaches a value $B_{0}$ (typically a few tens or hundreds of gauss) at which the scattering length between the atoms diverges. Let us first discuss the interaction of two atoms near a Feshbach resonance. The simplest way to describe it is to consider two coupled collisional channels: a background open channel with a certain scattering length $a_{\mathrm{bg}}$ coupled to a bare bound state lying in a closed channel. The bare bound state's energy $E_\mathrm{b}=\delta\mu_{B}(B-B_{0})$ with respect to the open channel varies linearly with $B$ by the Zeeman effect through the magnetic moment difference $\delta\mu_{B}$~\footnote{Here, $\delta \mu_B$ refers to the difference $\delta \mu_B =\mu_\mathrm{atoms} - \mu_\mathrm{molecule}$ between the respective magnetic moments $\mu_\mathrm{atoms}$ and $\mu_\mathrm{molecule}$ of separated atoms and the molecular bare bound state. For further details, see Ref. \cite{chin2010}.}. Assuming for simplicity contact interactions in both channels, one obtains an effective pair-interaction given by a contact interaction with a momentum-dependent scattering length~\cite{pricoupenko2010a,naidon2025}
\begin{equation}\label{eq:2bodyfeshbach}
    a(k) = a_{\mathrm{bg}} - \frac{1}{R^\star(\frac{m}{\hbar^2}E_\mathrm{b}-k^2)}.
\end{equation}
where the length $R^\star$ characterizes the strength of the coupling between the two channels. We emphasize that the terminology here refers to a two-channel description of the subsystem of \textit{two} bosons, not the full \textit{three} boson system. This interaction supports two bound states (see black curves in the upper panel of Fig.~\ref{fig:3d1}): one (Feshbach dimer) that dissociates at the Feshbach resonance ($e_{\mathrm{b}}=0$) and a lower one (deep dimer) which induces a continuum into which three-body resonances can decay.

We employ this model to obtain the exact solutions for the three-body system via the Skorniakov -- Ter-Martirosian equation~\cite{skorniakov1957}, regularizing the Thomas collapse~\cite{thomas1935} by a momentum cutoff $P_{\mathrm{cut}}=4{R^\star}^{-1}$. Using the complex-scaling method~\cite{moiseyev1998}, we can extract both real part (position) and imaginary part (width) of the lowest three-body resonance energy. The results are displayed in Fig.~\ref{fig:3d1}: the upper panel shows the resonance position and its width as a red curve with a shaded area. The lower panel solely depicts the width in log-scale. Both panels show the behavior as a function of the magnetic field $B$. One can clearly identify a zero of the resonance width, i.e. a point where the resonance turns into a stabilized bound state in the continuum (green cross). The existence and single-resonance parametric nature of this BIC is again confirmed by solving an effective two-channel \textit{three-body} model of the form of Eq.~\eqref{eq:2chnl}, obtained using an adiabatic hyperspherical representation (see Methods). The width of the resonance in this two-channel model is obtained from Eq.~\eqref{eq:gammaisolatedresonance} and shown in the lower panel of Fig.~\ref{fig:3d1}. It features a zero, in qualitative agreement with the exact calculation. In both cases the zero is located in an experimentally feasible regime for the magnetic field.

\begin{figure}[htbp]
    \centering
    \includegraphics[width=\columnwidth]{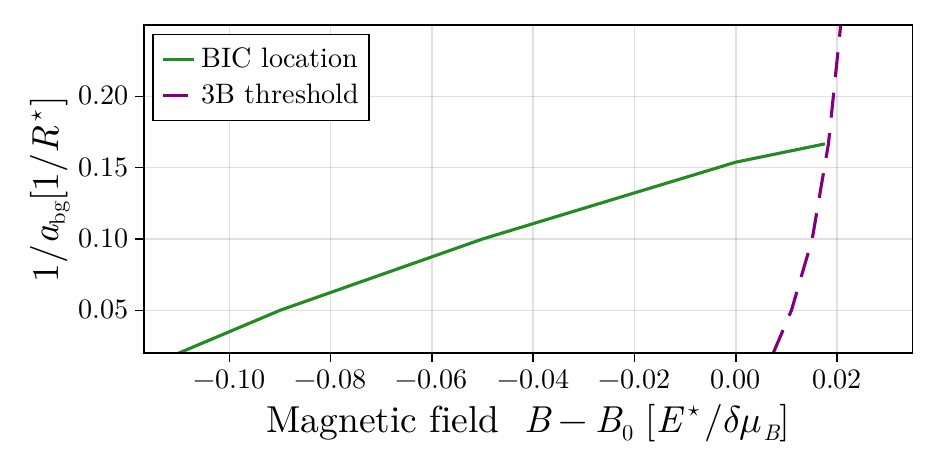}
    \caption{\textbf{Robustness of BIC existence against variation of the background scattering length.} BIC-location (green line) for three identical bosons in 3D, in the parameter space of the inverse background scattering length $1/a_\mathrm{bg}$ and the magnetic field $B$. The scattering length is given in units of $R^\star$. Additionally, the three-body binding threshold, $E_\mathrm{R}^{(3)}=0$, is displayed (dashed purple line).}
    \label{fig:3d2}
\end{figure}

Due to its parametric nature, the existence of the BIC is robust over a finite range of the system parameters. In this range, varying $a_{\mathrm{bg}}$ shifts the BIC location without destroying it. This feature is displayed in Fig.~\ref{fig:3d2} which tracks the BIC location in the parameter space spanned by $1/a_{\mathrm{bg}}$ and $B$. This demonstrates that a BIC can be found by tuning of several variables and for several values of the background scattering length.

\section{Discussion}
To summarize, we have presented a general theory for the stabilization of few-body resonances to bound states in a continuum. Based on a two-channel model, we have traced back its origin to a transition element of a single isolated resonance in a single continuum. This transition element can be made to vanish by variation of several system parameters. It is this tunability that allows us to classify these BICs to be of the single-resonance parametric type. Consequently, they can persist under variations of these parameters over finite ranges. While the underlying mechanism is general, the precise conditions for the occurrence of a vanishing width depend on the details of the specific system. Moreover, we have showcased the existence of three-body BICs in two different systems, demonstrating the generality of our stabilization mechanism. Most notably, we have shown that stabilization occurs in a system of three identical bosons in 3D, a system highly relevant to Efimov physics. Crucially, here the relevant tuning parameter to achieve stabilization and existence of a three-body BIC is directly related to an external magnetic field, a variable commonly tuned in cold-atom experiments. In both cases the two-channel model has provided a good qualitative description of the existence of BICs and their approximate location within the explored parameter ranges.

We expect that the present demonstration of few-body BICs will stimulate future studies to calculate the precise parameter values at which BICs occur in various systems. Finally, let us mention that the width of a three-body resonance, especially in the Efimov scenario, can be interpreted as the imaginary part of the three-body parameter. Its real part has been the subject of many studies, e.g. on the van-der-Waals universality~\cite{wang2012,endo2014}. It would be interesting to investigate whether the imaginary part and hence the stabilization of three-body resonances can show a similar universal behavior.

\section{Methods}\label{app:endmatter}

\subsection{Born-Oppenheimer picture}
In this section, we outline how to use the Born–Oppenheimer approximation to construct two effective three-body channels, and to extract the corresponding quantities required for evaluating the resonance width via Eq.~\eqref{eq:gammaisolatedresonance}.

For the one-dimensional, mass-imbalanced three-body system introduced before, we define $z$ as the relative distance between the two identical bosons and $Z$ as the distance of the third particle from their center of mass. In the three-body center-of-mass frame, the system is fully described by these two relative coordinates, with reduced masses $\mu_{\mathrm{bb}}$ and $\mu_{\mathrm{x},\mathrm{bb}}$ for the boson pair and the third particle relative to the pair, respectively.

In the first step, we neglect the dynamics of the two bosons in the dynamics of the light particle, i.e. the relative distance $z$ is considered to be frozen, and it enters only parametrically. This leads to the light-particle equation
\begin{equation}
    \left[-\frac{\hbar^2}{2\mu_{\mathrm{x},\mathrm{bb}}} \frac{\partial^2}{\partial Z^2} + V(Z_+) + V(Z_-) - U_\mathcal{C}(z)\right]\phi_\mathcal{C}(Z;z) = 0,
\end{equation}
with $\langle \phi_\mathcal{C}|\phi_\mathcal{C'}\rangle = \delta_{\mathcal{C},\mathcal{C'}}$ and where $Z_\pm = Z \pm z/2$ are the distances from the light particle to each of the two bosons, respectively. The parametric dependence on $z$ is highlighted by the semicolon notation in $\phi(Z;z)$. The eigenvalues $U_\mathcal{C}(z)$ then define, as a function of $z$, the effective potential curves for the heavy-particle dynamics. Each potential curve can be associated with a channel for the three-body system. In principle, this procedure yields infinitely many channels.

As a second step, we retain only the two lowest-energy states $|\phi_\mathcal{C}\rangle$, $\mathcal{C}\in\{\mathrm{o},\mathrm{c}\}$ and assign them to the open and closed channels of the three-body system. This reduces the full three-body problem to its two-channel form of Eq.~\eqref{eq:2chnl}, and the governing Hamiltonian reads
\begin{align}\label{eq:HCCnew}
    H_{\mathcal{C},\mathcal{C'}} =&~ \delta_{\mathcal{C},\mathcal{C'}}\left[-\frac{\hbar^2}{2\mu_{\mathrm{bb}}} \frac{\partial^2}{\partial z^2} + U_\mathcal{C}(z)\right] \nonumber \\
    &~-\frac{\hbar^2}{2\mu_{\mathrm{bb}}} \left[2\langle \phi_\mathcal{C}|\frac{\partial}{\partial z}|\phi_\mathcal{C'}\rangle\frac{\partial}{\partial z} + \langle \phi_\mathcal{C}|\frac{\partial^2}{\partial z^2}|\phi_\mathcal{C'}\rangle \right].
\end{align}
The closed-channel bare bound state $\Phi_{\mathrm{c},n}$ with energy $E_{\mathrm{c},n}$ is obtained from solving the corresponding Schr\"odinger equation for $\mathcal{C} = \mathcal{C'} = \mathrm{c}$, and the open-channel scattering state $\Phi_{\mathrm{o},k}$ is obtained from solving the same equation for $\mathcal{C} = \mathcal{C'} = \mathrm{o}$ with $ k^2 = 2\mu_{\mathrm{bb}} E_{\mathrm{c},n} /\hbar^2$. Finally, considering the off-diagonal coupling terms $H_{\mathrm{o,c}}$, the closed-channel bare bound state becomes a resonance, whose width can be estimated via Eq.~\eqref{eq:gammaisolatedresonance}.

\subsection{Hyperspherical picture}%\label{app:hyperspherical}
In this section, it is shown how the three-boson problem near a two-body Feshbach resonance can be cast into the two-channel form of Eq.~\eqref{eq:2chnl}.

Since the pairwise interaction $V(r)$ acts within two \emph{two-body} channels (open and closed), it is written as \begin{equation}
V\equiv\frac{\hbar^{2}}{m}\left(\begin{array}{cc}
v^{\mathrm{(o,o)}} & v^{\mathrm{(o,c)}}\\
v^{\mathrm{(c,o)}} & v^{\mathrm{(c,c)}}
\end{array}\right)\label{eq:two-channel-pairwise-potential}
\end{equation}
where $v^{\mathrm{(o,o)}}$ and $v^{\mathrm{(c,c)}}$ represent the pairwise potentials in the respective open and closed channel, while $v^{\mathrm{(o,c)}}=v^{\mathrm{(c,o)}}$ denote the interchannel couplings. For simplicity, these interactions are taken to be of the contact form, which can be implemented by the use of pseudopotentials,
\begin{align}
v^{(i,j)}(r) & \equiv4\pi a^{(i,j)}\delta^{3}(\bm{r})\frac{\partial}{\partial r}\left(\cdot\right)\label{eq:pseudopotentials}
\end{align}
with the scattering lengths $a^{(i,j)}$.

One then retrieves the momentum-dependent scattering length of Eq.~\eqref{eq:2bodyfeshbach}, with $a_{\mathrm{bg}}=a^{\mathrm{(o,o)}}$, in the limit when the energy difference $\Delta$ between the two channels goes to infinity while $a^{\mathrm{(c,c)}}$ goes to zero to maintain the energy of the bare bound state in the closed channel,
\begin{equation}
\Delta-\frac{\hbar^{2}}{m[a^{\mathrm{(c,c)}}]^{2}}\to\frac{\hbar^{2}}{m}e_{\mathrm{b}},\label{eq:limit1}
\end{equation}
and $a^{\mathrm{(o,c)}}=a^{\mathrm{(c,o)}}$ goes to zero to maintain the ratio 
\begin{equation}
\frac{2[a^{\mathrm{(o,c)}}]^{2}}{[a^{\mathrm{(c,c)}}]^{3}}\to\frac{1}{R^{\star}}.\label{eq:limit2}
\end{equation}

\begin{figure}
    \centering
    \includegraphics[width=\linewidth]{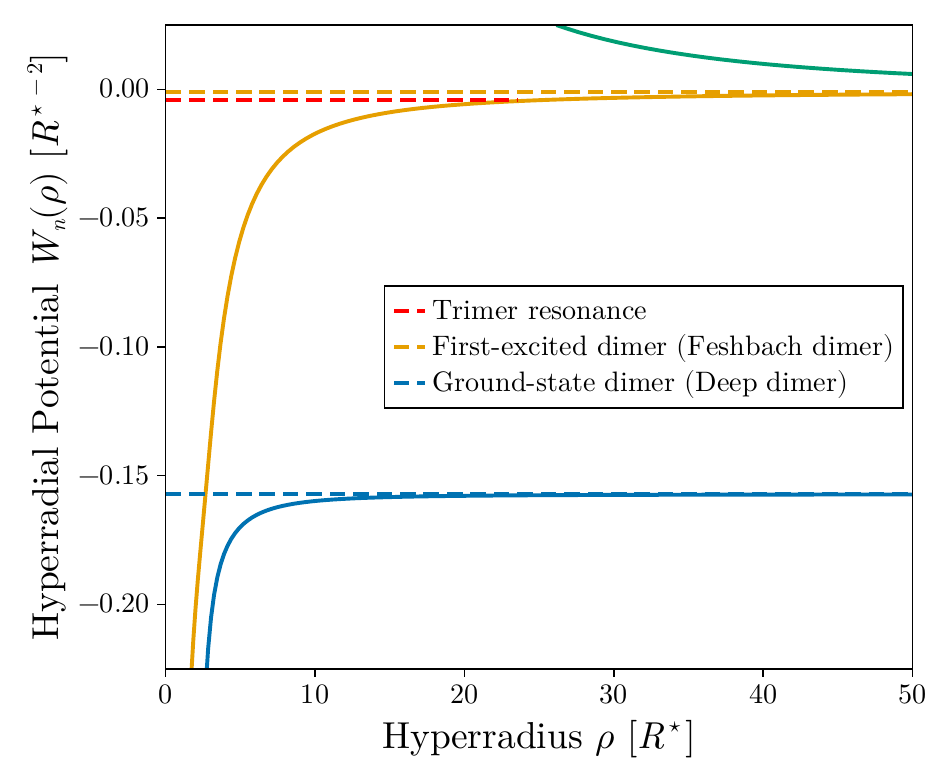}
    \caption{\textbf{Hyperradial potential curves with dimer and trimer spectrum.} Hyperradial potentials $W_{n}(\rho)$ for the two-channel contact model with $e_{\mathrm{b}}=-0.1R^{\star-2}$ and $a_{\mathrm{bg}}=20R^{\star}$. The energies of the dimer and trimer states are indicated by horizontal dashed lines.}
    \label{fig:hyperradial-potentials}
\end{figure}

To describe the three-body problem in this setting, one may consider at least two components for the three-body wave function,
\begin{equation}
\Psi\equiv\left(\begin{array}{c}
\Psi^{(\mathrm{o})}\\
\Psi^{(\mathrm{c})}
\end{array}\right),\label{eq:three-body-wavefunction}
\end{equation}
where $\Psi^{\mathrm{(o)}}$ corresponds to a state in which all pairs interact in the open channel, whereas $\Psi^{\mathrm{(c)}}$ is a state where one of the pairs interact in the closed channel. These two components, however, do not constitute the three-body open and closed channels we are looking for. To obtain such three-body channels, one can start from the adiabatic hyperspherical representation~\citep{nielsen2001a}. The three-body wave function is first expressed in hyperspherical coordinates ($\rho,\Omega$), where $\rho$ designates the hyperradius and $\Omega$ the hyperangles, and is expanded as
\begin{equation}
\Psi(\rho,\Omega)=\rho^{-5/2}\sum_{n=1}^{\infty}f_{n}(\rho)\Phi_{n}(\Omega;\rho)\label{eq:expansion}
\end{equation}
on an adiabatic basis of two-component hyperangular wave functions
$\Phi_{n}(\Omega;\rho)\equiv\left(\begin{array}{c}
\Phi_{n}^{(\mathrm{o})}(\Omega;\rho)\\
\Phi_{n}^{(\mathrm{c})}(\Omega;\rho)
\end{array}\right)$, which are chosen to be eigenstates of the interacting system at a fixed hyperradius $\rho$, with eigenvalues $\gamma_{n}(\rho)$. Using the two-channel contact interaction Eqs.~(\ref{eq:two-channel-pairwise-potential}-\ref{eq:pseudopotentials}) in the large-separation limit Eqs.~(\ref{eq:limit1}-\ref{eq:limit2}), one finds the following transcendental equations for $\gamma_{n}$~\citep{sorensen2013}:
\begin{equation}
-\gamma_{n}\cos\left(\frac{\pi}{2}\gamma_{n}\right)+\frac{8}{\sqrt{3}}\sin\left(\frac{\pi}{6}\gamma_{n}\right)+\frac{\rho}{a(\gamma_{n}/\rho)}\sin\left(\frac{\pi}{2}\gamma_{n}\right)=0\label{eq:transcendentalEq}
\end{equation}
One can then write an exact set of coupled equations for the remaining
hyperradial functions $f_{n}(\rho)$:
\begin{multline}
\left(-\frac{d^{2}}{d\rho^{2}}+W_{n}(\rho)+\kappa^{2}\right)f_{n}(\rho)+\sum_{p=1}^{\infty}H_{n,p}f_{p}(\rho)=0\label{eq:hyperradialEqs}
\end{multline}
with
\begin{align}
H_{n,p} & \equiv-P_{n,p}(\rho)\frac{d}{d\rho}-\frac{d}{d\rho}\left(P_{n,p}(\rho)\cdot\right)+Q_{n,p}(\rho)\cdot\label{eq:Hnp}\\
P_{n,p}(\rho) & \equiv\langle\Phi_{n}\vert\frac{d\Phi_{p}}{d\rho}\rangle\label{eq:P-definition}\\
Q_{n,p}(\rho) & \equiv\langle\frac{d\Phi_{n}}{d\rho}\vert\frac{d\Phi_{p}}{d\rho}\rangle\label{eq:Qtilde-definition}
\end{align}
and where $W_{n}(\rho)$ designate the hyperradial potentials $\frac{\gamma_{n}^{2}(\rho)-1/4}{\rho^{2}}$ obtained from solving Eq.~\eqref{eq:transcendentalEq}. The lowest three hyperradial potentials are represented in Fig.~\ref{fig:hyperradial-potentials}, at a particular value of $e_{\mathrm{b}}=-0.1R^{\star-2}$ and $a_{\mathrm{bg}}=20R^{\star}$. The lowest curve asymptotes to the energy of the ground-state dimer (the ``deep dimer''), while the next curve asymptotes to the energy of the first-excited dimer (the ``Feshbach dimer''). A trimer bound state lying in the upper channel can therefore decay into an atom scattering with a deep dimer in the lower channel, and thus turn into a three-body resonance. Restricting our analysis to these two lowest channels, one thus obtains an effective Hamiltonian of the form of Eq.~\eqref{eq:2chnl}, that captures the essence of the resonance. These two channels (with indices $\mathrm{o}$, $\mathrm{c}$ corresponding to $n=1,2$) now constitute the \emph{three-body} open and closed channels, not to be confused with the \emph{two-body} open and closed channels (with superscripts $^{\mathrm{(o)}}$ and $^{\mathrm{(c)}}$) describing the two-body interactions.

Although the contact interaction model given by Eqs.~(\ref{eq:two-channel-pairwise-potential}-\ref{eq:limit2}) is well defined for two particles, it has long been known to lead to the so-called Thomas collapse of three particles~\citep{thomas1935,naidon2017}. In our exact STM calculations, the problem is cured by imposing a cutoff in the dimer-particle relative momentum. Here, the collapse manifests itself by excessive attraction in the short-distance region of the hyperradial potentials. It can be cured by imposing a short-distance repulsion, which we choose to be a hard wall located at some hyperradius $\rho_{\mathrm{min}}$ in each channel. The values of $\rho_{\mathrm{min}}$ are adjusted to reproduce the energy $E_{\mathrm{o},1}$ of the bound trimer below the deep dimer, and the energy $E_{\mathrm{c},1}$ of the trimer resonance below the Feshbach dimer, as found in our exact STM calculations. At that stage, one can determine both the bare three-body bound state $\vert f_{\mathrm{c},1}\rangle$ in the closed channel, and the corresponding scattering state $\vert f_{\mathrm{o},k}\rangle$ in the open channel at the same energy $E_{\mathrm{c},1}$. Finally, by considering the coupling terms $P_{\mathrm{o,c}}$, $Q_{\mathrm{o,c}}$, one can estimate the resonance width within this two-channel approximation, based on Eq.~\eqref{eq:gammaisolatedresonance}.

\section{Data Availability}
The data that support the findings of this study are available from the authors upon reasonable request.

\section{Code Availability}
All code that supports the findings of this study are available from the authors upon reasonable request.

%\bibliography{Stabilization}

%\begin{acknowledgements}
\section{Acknowledgments}
L.~H. is supported by the RIKEN special postdoctoral researcher program. P.~N. acknowledges support from the JSPS Grants-in-Aid for Scientific Research on Innovative Areas (No.JP23K03292).
%\end{acknowledgements}

\section{Author contributions}
All authors contributed equally to this work.

\end{document}